\documentstyle[epsfig,11pt]{article}
\title{\bf Forward pomeron propagator in the external field of the nucleus}
\author{M.A.Braun, A.Tarasov  \\
S.Peterburg State University, Russia}
\begin{document}
\maketitle
\input epsf
\def\beq{\begin{equation}}
\def\eeq{\end{equation}}
\def\phid{\Phi^{\dagger}}
\def\pd{\partial}
\medskip

{\bf Abstract}

It is shown by numerical calculations that the convoluted forward
pomeron propagator in the external field created by a solution of the
Balitski-Kovchegov equation in the nuclear matter vanishes at high
rapidities. This may open a possibility to apply the perturbative approach
for the calculation of pomeron loops.

\section{Introduction}
In the QCD, in the limit of large number of colours,
strong interaction at high energies is mediated by the exchange of
BFKL pomerons, which interact via their splitting and fusion.
In the quasi-classical
approximation  for photon (hadron)-nucleus scattering the relevant
tree (fan) diagrams
are summed by the well-known Balitski-Kovchegov (BK) evolution equation
 ~\cite{bal,kov,bra1}. For nucleus-nucleus scattering appropriate
 quasi-classical equations were derived in ~\cite{bra2,bra3}.
In both cases pomeron loops were neglected.
This  approximation can be
justified if the parameter $\gamma=\lambda\exp{\Delta y}$ is small,
with $y$ the rapidity and $\Delta$ and $\lambda$ the pomeron intercept
and triple pomeron coupling.
Then for a large nuclear target, such that $A^{1/3}\gamma\sim 1$,
the tree diagrams indeed give the dominant contribution and loops
can be dropped. However with the growth of $y$ the loop
contribution becomes not small and this approximation breaks down.

Direct calculation of the loop contribution seems to
be a formidable task  for the non-local BFKL pomeron.
Simplest loops have been studied
in several papers for purely hadronic scattering ~\cite{peschansky,
bartels, bra4}. In particular in ~\cite{bra4} it has been found
that pomeron loops become essential already at rapidities of the
order 10$\div 15$. They shift the position of the pomeron pole
to the complex plane and thus lead to oscillations in cross-sections.
However with the growth of energy loop contributions begin to dominate
and one needs to sum all of them. There have been many attempts to do this
in the framework of the so-called reaction-diffusion formulation of the
QCD dynmaics and the following correspondence with the statistical approach
~\cite{iancu,AHM,levlub,lev,marquet,levmil} (see also a review
~\cite{soyez} and references therein).
Unfortunately concrete results could be  obtained only
with very crude approximations  for the basic BFKL interaction and the
stochastical noise in the statistical formulation.
The conclusions of different groups are incomplete and contradictory.
So in ~\cite{levmil} it was found that the geometric scaling following
from the BK equation was preserved with loops taken into account,
although going to the black disc limit was much slower.
On the contrary in papers based on the analogy with statistical phyiscs
(see ~\cite{marquet,soyez}) it was argued that the BK scaling was changed to the so called
diffusive scaling (with an extra $\sqrt{y}$ in the denominator of the
argument) but the speed of achieving the black disk limit was essentially
unchanged.

In our previous study of pomeron loops ~\cite{BT}
we considered a much simpler model with the local supercritical pomeron
in the Regge-Gribov formalism.
Instead of trying to solve
the model for the purely hadronic scattering we considered
the hadron-nucleus scattering and propagation of the pomeron inside
the heavy nucleus target. Moreover to avoid using numerical
solution of the tree diagrams contribution with
diffusion in the impact parameter, we concentrated on the case of
a constant nuclear density which allowed to start with the known
analytical solutions. We have found that the
nuclear surrounding  transforms the pomeron from the supercritical
one with intercept $\epsilon>0$ to a subcritical one with the intercept
$-\epsilon$. Then Regge cuts, corresponding to loop diagrams, start at
branch points located to the left of the pomeron pole and their contribution
is subdominant at high energies. As a result the theory aquired the
properties similar to the Regge-Gribov with a subcritical pomeron
and allows for application of the perturbation theory. In~\cite{BT}
we expressed our hopes that a similar phenomenon might occur in the
QCD with BFKL pomerons.

In this note we demonstrate that such hopes are possibly founded.
We consider the pomeron propagator in the external pomeron field created
inside the nucleus and give arguments that, similar to the local
Regge-Gribov case,
it vanishes at large rapidity distances. We stress that at present
we are unable to give the full proof for this behaviour. Our study is
based on numerical calculations.
This makes us to choose  a relatively small
subset of initial conditions out of the complete set
necessary for the study of the pomeron propagator.
Moreover, due to technical difficulties, in this note we restricy ourselves
to the much simpler pomeron propagator in the forward direction.
Our numerical results show that, with the chosen set of initial conditions,
this forward  propagator vanishes at large rapidity distances.
This result  is insufficient for the study of loops, where non-forward propagator
are involved. However it  can be applied for
double inclusive cross-sections in the nucleus-nucleus scattering, in which
only forward propagators are important.

\section{Main equation}
\begin{figure}
\hspace*{4 cm}
\epsfig{file=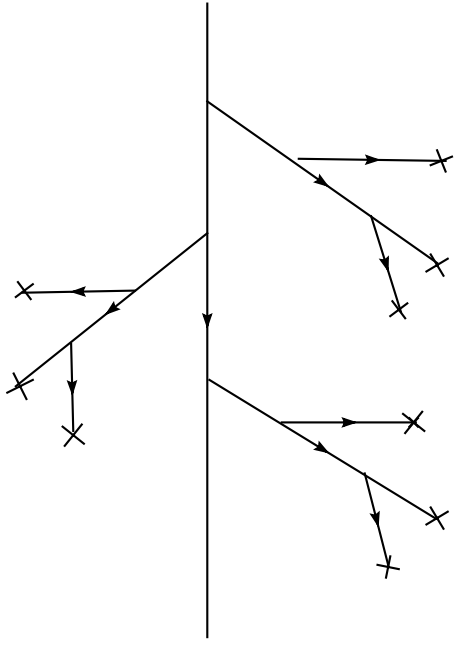,width=6 cm}
\caption{Diagrams summed into the pomeron prpagator in the nuclear matter}
\label{fig1}
\end{figure}

We are going to  study the behaviour of the BFKL pomeron propagator
in the external field, generated by a solution of the
BK equation.
We consider the simplified case of the nuclear matter, when the
dependence on the impact parameter $b$ is absent.
This propagator corresponds to a sum of diagrams shown in Fig. \ref{fig1}.

As mentioned, in this note we restrict ourself to the forward propagator,
$P(y,x,x')$
when the transferred momentum is zero. It depends on rapidity $y$
and two 2-dimensional coordinate
vectors $x$ and $x'$ corresponding to  the initial and
final distances between the reggeized gluons in the pomeron.
At $y=0$ we have $P(y,x,x')=\nabla^{-4}\delta^2(x-x')$.
Since the study is only possible numerically,
to avoid using this singular initial condition,
we shall consider a convolution of $P(y,x,x')$ with an arbitrary
initial function $\nabla^4\psi(x)$
\beq
P(y,x)=\int d^2x'P(y,x,x')\nabla^4\psi(x').
\eeq
This convolution satisfies the same equation as the propagator itself
but at $y=0$ we have
\beq
P(y=0,x)=\psi(x).
\label{ini}
\eeq
Obviously properties of the propagator can be studied taking a full
set of functions $\psi(x)$.

The equation for $P(y,x)$ can be conveniently obtained from the
BK equation for the sum of fan diagrams. In the forward
direction this sum $\Phi(y,x)$ satisfies a non-linear eqiation
\beq
\frac{\partial\Phi(y,x)}{\partial y}=
\frac{\bar{\alpha}}{2\pi}\int d^2x_1\frac{x^2}{x_1^2x_2^2}
\Big(\Phi(y,x_1)+\Phi(y,x_2)-\Phi(y,x)-\Phi(y,x_1)\Phi(y,x_2)\Big),
\label{bk}
\eeq
where standardly
\[\bar{\alpha}=\frac{\alpha_sN_c}{\pi}.\]
The equation for the convoluted propagator $P(y,x)$ in the
presence of nuclear medium is obtained when one of the $\Phi$ in
the non-linear term in
(\ref{bk}) is substituted by a particular solution  of (\ref{bk})
with a given boundary condition. The equation thus obtained is
\beq
\frac{\partial P(y,x)}{\partial y}=
\frac{\bar{\alpha}}{2\pi}\int d^2x_1\frac{x^2}{x_1^2x_2^2}
\Big(P(y,x_1)+P(y,x_2)-P(y,x)-2\Phi(y,x_1)P(y,x_2)\Big).
\label{eq1}
\eeq
Note that the initial condition for $\Phi(y=0,x)=\Phi_0(x)$ is fixed by the
properties of the nuclear medium, whereas the initial condition (\ref{ini})
for $P(y,x)$ is arbitrary, since we we are interested in the propagator in a
given nuclear surrounding.

Equation (\ref{eq1}) is a linear equation for $P(y,x)$ in contrast to the
BK equation. At $y\to\infty$ $\Phi(y,x)\to 1$
independent of the chosen initial condition. One may think that at
$y\to\infty$ the behaviour of $P(y,x)$ can be derived from the asymptotic
equation
\[
\frac{\partial P(y,x)}{\partial y}\Big|_{y\to\infty}=
\frac{\bar{\alpha}}{2\pi}\int d^2x_1\frac{x^2}{x_1^2x_2^2}
\Big(P(y,x_1)+P(y,x_2)-P(y,x)-2P(y,x_2)\Big)\]\beq=
-P(y,x)\frac{\bar{\alpha}}{2\pi}\int d^2x_1\frac{x^2}{x_1^2x_2^2}.
\label{eq11}
\eeq
However the integral on the right-hand side has become divergent
(although it converges at finite $y$). This means that the limit
$y\to\infty$ is more delicate and cannot be taken under the sign of integral
over $x_1$. And indeed we shall see by numerical calculation that
the behaviour of the solution at $y\to\infty$ is not solely determined
by the limiting value of $\Phi(y,x)$ but depends on its behaviour
at finite $y$.

For numerical studies  both the BK equation and
linear equation (\ref{eq1}) in the momentum space are more convenient.
Introducing
\beq
\phi(y,x)=\frac{\Phi(y,x)}{x^2},\ \ p(y,x)=\frac{P(y,x)}{x^2}
\eeq
and then passing to the momentum space we obtain the following equations for
$\phi(y,k)$ and $p(y,k)$
\beq
\frac{\partial\phi(y,k)}{\partial y}=
-\bar{\alpha}\Big(H_{BFKL}\phi(y,k)+\phi^2(y,k)\Big)
\label{bkmom}
\eeq
and
\beq
\frac{\partial p(y,k)}{\partial y}=
-\bar{\alpha}\Big(H_{BFKL}+2\phi(y,k)\Big)p(y,k),
\label{eq1mom}
\eeq
where
\beq
H_{BFKL}=\ln k^2+\ln x^2-2(\psi(1)+\ln 2).
\eeq
To study the behaviour of the propagator in the external field $\phi$
one has to solve this pair of equations with the initial conditions
\beq
\phi(y,k)_{y=0}=\phi_0(k),\ \ p(y,k)_{y=0}=p_0(k),
\label{inimom}
\eeq
with some fixed $\phi_0$ and for a complete set of function $p_0(k)$.

\section{Numerical studies}
\begin{figure}
\hspace*{2 cm}
\epsfig{file=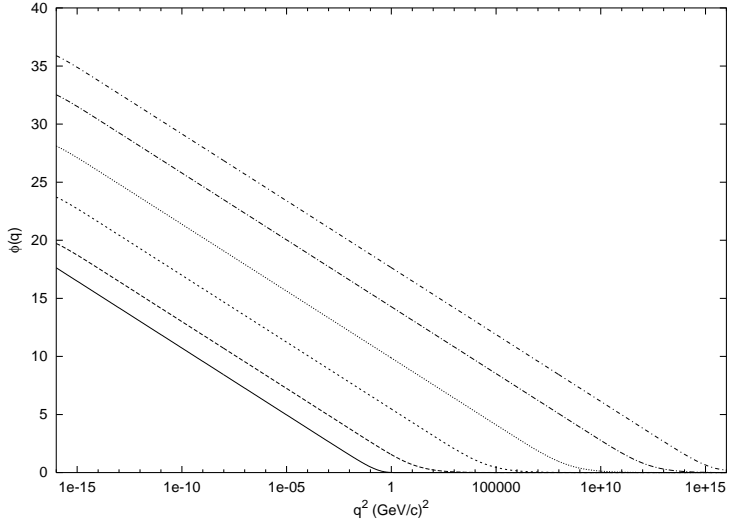,width=10 cm}
\caption{Solution $\phi(y,q)$ of Eq. (\ref{bkmom}) as a function of $q^2$
for different of $Y=\bar{\alpha}y$.
Curves from bottom to top correspond to $Y=0,2,4,6,8$ and 10.}
\label{fig2}
\end{figure}

We have set up a program which simultaneously solves the
pair of equations (\ref{bkmom}) and (\ref{eq1mom}) for a given pair
of initial conditions (\ref{inimom}). For the BK evolution we
have fixed the initial condition as
\beq
       \phi_0(k)=-(1/2){\rm Ei}(-k^2/0.3657)
\eeq
used in our previous calculations. The behaviour of $\phi(k)$
with $k^2$ at different values of the scaled rapidity
$Y=\bar{\alpha}y=2,4,6,8$ and 10
is shown in Fig. \ref{fig2}.
(Note that the maximal value of the scaled rapidity $Y=10$ corresponds
to the natural rapidity of order 50).

For the BFKL evolution in the external field $\phi$, in the first run (A),
we have taken the same form of the initial condition but with
a variable slope
\beq
       p_0(k)=-(1/2){\rm Ei}(-k^2/a).
\label{ini1}
\eeq
We have performed calculations for $a=0.2,\  0.6,\ 1.0,\ 1.4$ and 1.8.
In the second run (B) the initial condition was taken with extra powers
of $k^2$.
\beq
       p_0(k)=-(1/2)k^{2n}{\rm Ei}(-k^2/0.3657)
\label{ini2}
\eeq
with $n=0,1,2,3$ and 4.

In all cases the behaviour of the solution $p(y,k)$ was found to be
universal. At large enough $y$ the solution becomes independent of $k^2$
up to a certain maximal $k^2_{max}(y)$, starting from which it goes to zero.
Roughly
\beq
p(y,k)\sim A(y)\theta(k^2_{max}-k^2).
\label{pyk}
\eeq
As $y$ grows $A(y)$ goes to zero  and $k^2_{max}(y)$ goes to infinity. .
So on the whole the solution vanishes as $y\to\infty$, its $x$ dependence
tending to $\delta^2(x)$.

\begin{figure}
\hspace*{2 cm}
\epsfig{file=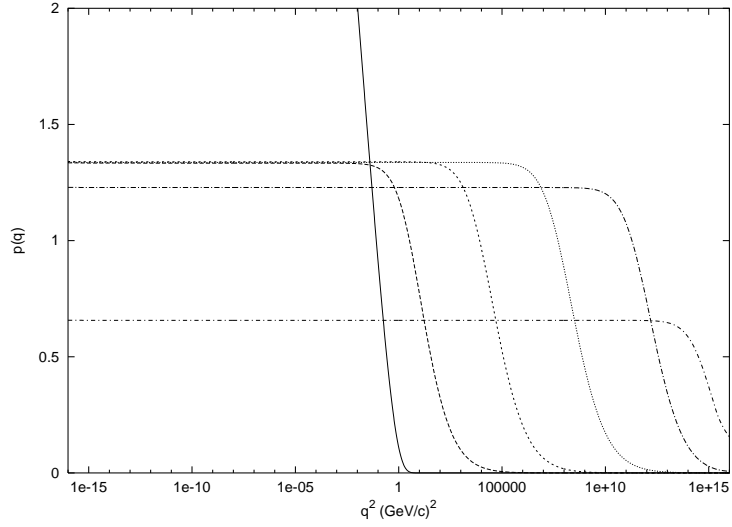,width=10 cm}
\caption{Solutions $p(y,q)$ of Eq. (\ref{eq1mom}) as a function of $q^2$
for different $Y=\bar{\alpha}y$ for run A with $a=1$. Curves which
start to fall at higher $q^2$
correspond to higher $Y=0,2,4,6,8$ and 10}
\label{fig3}
\end{figure}

\begin{figure}
\hspace*{2 cm}
\epsfig{file=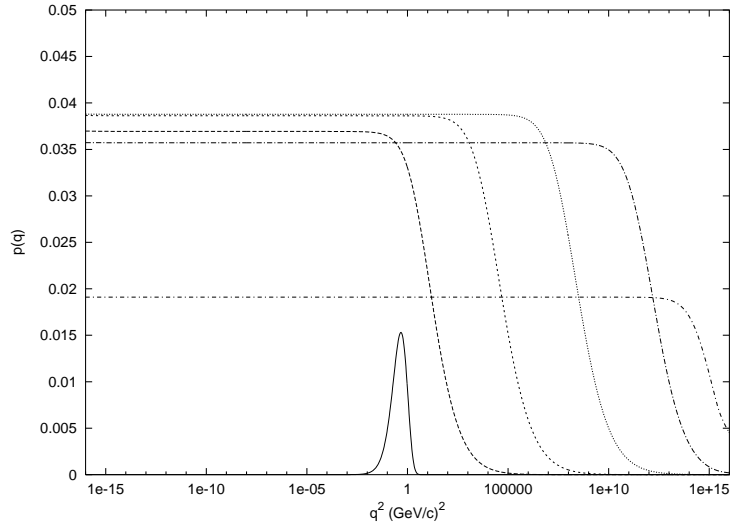,width=10 cm}
\caption{Solutions $p(y,q)$ of Eq. (\ref{eq1mom}) as a function of $q^2$
for different $Y=\bar{\alpha}y$ for run B with $n=2$. Curves which start to fall
at higher $q^2$ correspond to higher $Y=0,2,4,6,8$ and 10}
\label{fig4}
\end{figure}

We illustrate this behaviour in Figs. \ref{fig3} and \ref{fig4}, in which we show
the solution $p(y,k)$ for run A with $a=1.0$ and run B with $n=2$ as a function
of $k^2$.
One observes that although the values of $p(y,k)$ for the two cases are
different, their behavior with $y$ is the same: they vanish as $y\to\infty$.

\begin{figure}
\hspace*{2 cm}
\epsfig{file=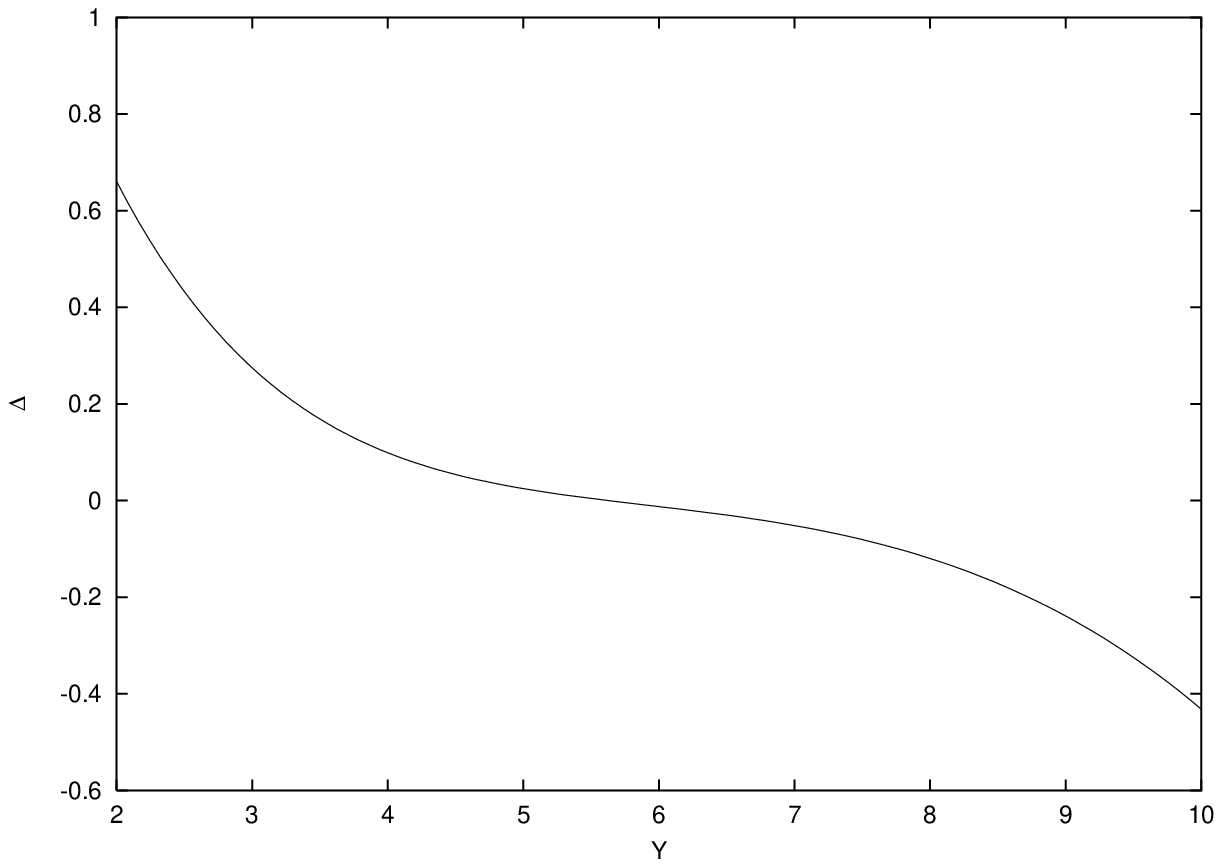,width=10 cm}
\caption{Slope $\Delta$ of the $Y$-dependence of the solutions $p(y,q)$
of Eq. (\ref{eq1mom}). The slope is the same for all initial conditions
(\ref{ini1}) and (\ref{ini2}).}
\label{fig5}
\end{figure}

This unversality is especially obvious if one calculates the slope
$\Delta(y,k)$ of the $y$-dependence of
$p(y,k)$ at fixed $k$ presenting
\beq
p(y,k)\propto e^{Y\Delta(y,k)}.
\eeq
It turns out that at $Y>1$ the slope $\Delta(y,k)$ is independent of $k$ and
identical for all considered cases (run A with all studied $a$ and run B
with all studied $n$). Its smooth behaviour with $Y$ is shown in Fig.
\ref{fig5}. One observes that starting from $Y=5$ the slope becomes negative
indicating that the solution goes to zero at $Y>>1$.

\begin{figure}
\hspace*{2 cm}
\epsfig{file=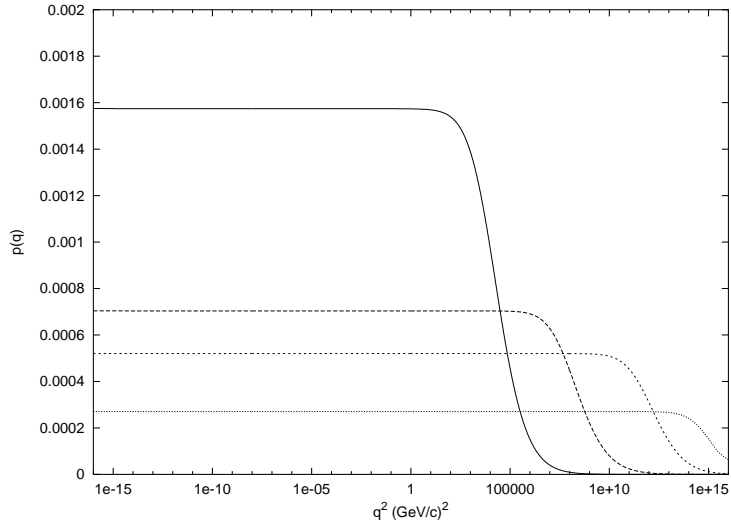,width=10 cm}
\caption{Solutions $p(y,q)$ of Eq. (\ref{eq1mom}) as a function of $q^2$
for different $Y=\bar{\alpha}y$ for the initial value taken at
$Y=3$ according to run A with $a=1$. Curves which
start to fall at higher $q^2$
correspond to higher $Y=4,6,8$ and 10}
\label{fig6}
\end{figure}

\begin{figure}
\hspace*{2 cm}
\epsfig{file=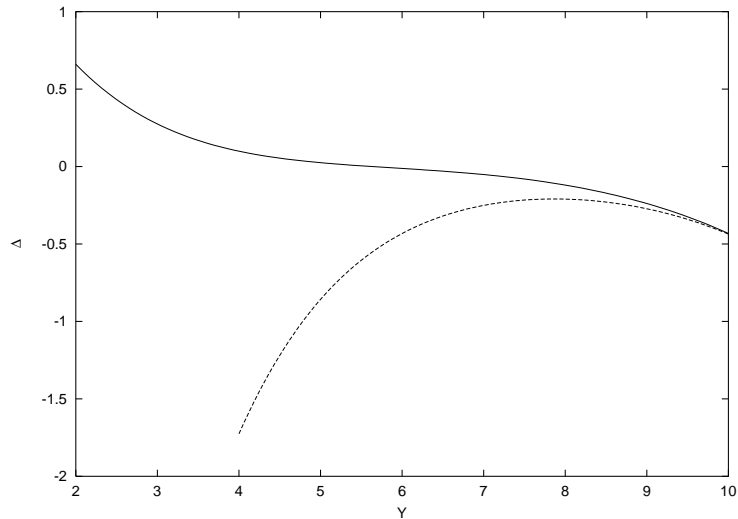,width=10 cm}
\caption{Slope $\Delta$ of the $Y$-dependence of the solutions $p(y,q)$
of Eq. (\ref{eq1mom}) with the initial condition put at $Y=0$ (upper curve)
and at $Y=3$ (lower curve) according to run A with $a=1$.}
\label{fig7}
\end{figure}

One has to take into account that in the external field $\phi(y,k)$
depending on rapidity the pomeron prpagator ceases to depend only on the
rapidity difference. Rather the initial and final rapidities become two
independent variables. To see what influence it has on the behaviour of
the propagator at large rapidities we varied the initial rapidity $y=y_0$
for the evolution of $p(y,k)$,  leaving unchanged the initial
rapidity $y=0$ for the evolution of $\phi(y,k)$,
which is the rapidity of the nucleus. One finds that although at initial
stages of evolution the behavior of $p(y,k)$ strongly depends on the
value of $y_0$, at higher rapidities this behaviour is essentially the same
for any $y_0$, namely the convoluted propagator goes down with rapidity
with the slope independent of $y_0$. This is illustrated in Figs. \ref{fig6}
and \ref{fig7} which show results for $y_0=3/\bar{\alpha}\ (Y_0=3)$
In Fig. \ref{fig6} we show the solution $p(y,k)$ for run A
with $a=1$.
One observes, that although absolute values of $p(y,k)$ are quite different
from the case $y_0=0$ shown in Fig \ref{fig3}, the behaviour with the
growth of rapidity is the same. It is especially clear from the values
for the slope $\Delta$ shown in Fig. \ref{fig7} together with those for
the case $y=0$ (Fig. \ref{fig5}). Again at the initial stage of the evolution
the behavior with $Y_0=3$ is quite different from that with $Y_0=0$.
However at higher rapidities the values for the slope become the
same.

\begin{figure}
\hspace*{2 cm}
\epsfig{file=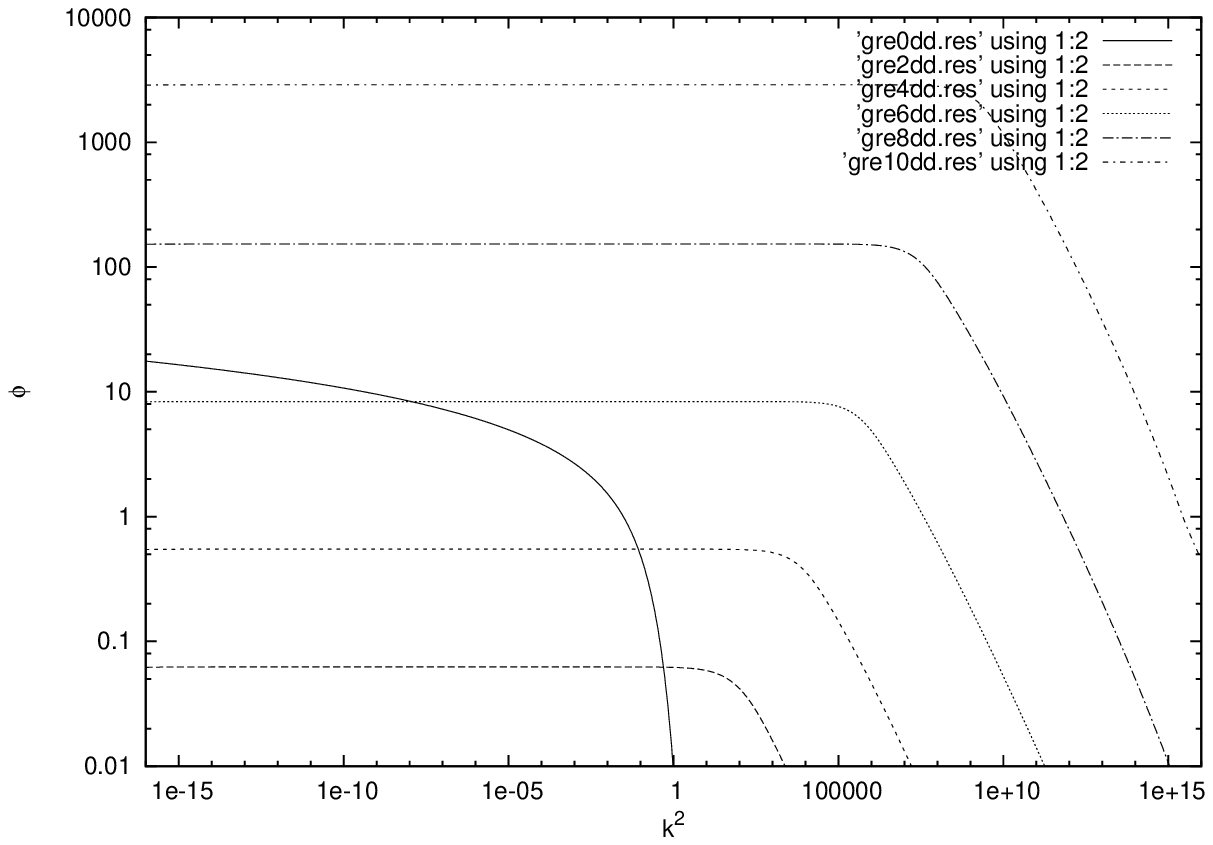,width=10 cm}
\caption{Solutions of Eq. (\ref{evoleq}) as a function of $k^2$ for different
$y$. Curves from bottom to top correspond to
$Y=\bar{\alpha}y=0,2,4,6,8$ and 10.}
\label{fig8}
\end{figure}

It is remarkable that this behaviour takes place only with $\phi$
given by the exact solution of the BK equation.
Taking an approximate form
\beq
\Phi(y,x)\simeq 1-e^{-Q^2(y)x^2},
\label{apprphi}
\eeq
where the "saturation momentum" $Q^2(y)\sim e^{2.05\bar{\alpha}y}$
we obtain an equation for $p(y,k)$ in the momentum space
\beq
\frac{\partial p(y,k)}{\partial y}=-\bar{\alpha}\Big[H_{BFKL}-
{\rm Ei}\Big(-\frac{k^2}{4Q^2(y)}\Big)\Big]
\label{evoleq}
\eeq
Taking for simplicity the initial condition $p_0(k)=\phi_0(k)$ at $y=0$
we get the solution shown in Fig. \ref{fig8}.
One observes that at large $y$ the solution acquires the same form
(\ref{pyk}) where however $A(y)$ grows with $y$:
\beq
A(y)\sim e^{1.4 Y}
\eeq
This implies that with the approximate form (\ref{apprphi}) of $\Phi$
the final solution $P(y,x)$ in the coordinate space
behaves in a singular manner at $y\to\infty$. Effectively
\beq
P(y,x)_{y\to\infty}\to e^{1.4 y}x^2\delta^2(x)
\eeq
and it is impossible to say that it vanishes in this limit.

\section{Conclusions}
We have studied numerically the BFKL pomeron forward propagator in
the external field created by the solution of the BK
equation in the nuclear matter. We have found that for more or less
arbitrary set of initial conditions the convoluted propagator
vanishes at large rapidities, its coordinate dependence tending to the
$\delta$-function.  This gives reasons
to believe that the forward propagator itsef vanishes at large rapidities
in the nuclear background. This result follows only with the field being the
exact solution of the BK equation.

Our results are obviously insufficient for the calculation of pomeron
loops, which requires the non-forward pomeron propagator. However
they can be directly applied to the study of double inclusive cross-section
for gluon jet production in nucleus-nucleus collisions. This highly
complicated problem is left for future investigation.

\section{Acknowledgments}
This work has been supported by grants RFFI 09-012-01327-a and RFFI-CERN
08-02-91004.

\end{document}